\documentstyle[twocolumn,floats,aps]{revtex}

\begin{document}
\draft

\twocolumn[\hsize\textwidth%
\columnwidth\hsize\csname@twocolumnfalse\endcsname

\title{\bf Many-Body Renormalization of Semiconductor Quantum Wire Excitons: Absorption, Gain, Binding, Unbinding, and Mott Transition}

\author{S. Das Sarma and D. W. Wang} 

\address{Department of Physics, University of Maryland, College Park, Maryland 20742-4111}

\date{\today}
\maketitle
\pagenumbering{arabic}

\begin{abstract} 
We consider theoretically the formation and stability of quasi-one dimensional
many-body excitons in GaAs quantum wire 
structures under external photoexcitation 
conditions by solving the dynamically screened Bethe-Salpeter equation 
for realistic Coulomb interaction. 
In agreement with several recent experimental findings the calculated excitonic
peak shows very weak carrier density dependence upto (and even above)
the Mott transition density, $n_c\sim3\times10^5$ cm$^{-1}$. Above $n_c$ we 
find considerable optical gain demonstrating compellingly the possibility
of one-dimensional quantum wire laser operation.
\end{abstract}

\pacs{PACS numbers: 78.55.-m; 71.35.Cc; 78.66.Fd; 73.20.Dx}
\vfill\eject
\vskip 1pc]
\narrowtext
An exciton, the bound Coulombic ("hydrogenic") state between an electron in 
the conduction band and a hole in the valence band, is an (extensively studied)
central concept in semiconductor physics. Recent interest has focused on low 
dimensional excitons in artificially structured semiconductor quantum well or 
wire systems where carrier confinement may substantially enhance the excitonic 
binding energy leading to novel optical phenomena. In this Letter we consider 
the formation, stability, and optical properties of one dimensional (1D) 
excitons in semiconductor quantum wires, a problem which has attracted a 
great deal of 
recent experimental [1-3] and theoretical [4-6] attention. Our motivation has 
been a number of recent puzzling experimental observations [1,2], which find 
the photoluminescence emitted from an initially photoexcited 
semiconductor quantum wire plasma to be peaked essentially at a 
\textit{constant} energy independent of the magnitude of the photoexcitation 
intensity. This is surprising because one expects a strongly density-dependent
"red shift" in the peak due to the exchange-correlation induced
band gap renormalization (BGR) (i.e. a 
density-dependent shrinkage of the fundamental 
band gap due to electron and hole self-energy corrections), which should vary 
strongly as a function of the photoexcited electron-hole density [7-9]. This 
striking lack of any dependence of the observed photoluminescence peak energy 
on the photoexcitation density has led to the suggestion [1,2] that the 
observed quantum wire photoluminescence may be arising entirely from an 
excitonic (as opposed to an electron-hole plasma (EHP)) 
recombination mechanism, and the effective 
excitonic energy is, for unknown reasons, a constant (as a 
function of carrier density) in 1D quantum wires. This, however, 
introduces a new puzzle because one expects the excitonic level to exhibit a 
"blue shift" (i.e. an increase) as a function of carrier density as the 
Coulomb interaction weakens due to screening by the finite carrier density 
leading to a diminished excitonic binding energy. Thus the only way to 
understand the experimental observation is to invoke a near exact cancelation
between the red-shift arising from the self-energy correction induced BGR
and the blue-shift arising from screening induced excitonic 
binding weakening. In this Letter, 
focusing on the photoexcited quasi-equilibrium
regime, we provide the first quantitative theory for 
this problem by solving the full many-body dynamical Bethe-Salpeter equation 
for 1D excitons. We include both self-energy renormalization and 
vertex correction (arising from the Coulomb interaction) on an equal footing 
under high photoexcitation conditions. We find that, in agreement with 
experimental observations, our calculated effective excitonic energy 
(indicating the luminescence peak frequency) remains essentially a constant 
(with an energy shift of less than 0.5 meV) 
as a function of 1D carrier density $n$ for $n<n_c\sim3\times10^5$
cm$^{-1}$ with the system making a Mott transition from an insulating exciton 
gas of bound electron-hole pairs ($n<n_c$) to an EHP 
($n>n_c$) at $n=n_c$. For $n>n_c$ we find strong optical gain in the calculated
absorption spectra.

For our results to be presented here 
we have considered quantum wire parameters [1]
corresponding to the T-junction structure of width 70 \AA\ in both 
transverse directions with only the lowest 1D 
subband occupied by the carriers. But our results and 
conclusions should be generically valid for arbitrary 1D 
quantum wire 
confinement (e.g. the V-groove wire of Ref. 2). The many-body exciton is given 
by the so-called Bethe-Salpeter equation [10] for the 2-particle Green's 
function 
which is shown diagrammatically in Fig. 1. The many-body diagrams 
shown in Fig. 1 correspond to a rather complex set of coupled 
non-linear integral 
equations which must be solved self-consistently with the bare interaction 
being the Coulomb interaction. These equations are notoriously difficult [10]
to solve without making drastic approximations. We use the parabolic band 
effective
mass approximation considering the highest valence and the lowest conduction 
band only. The simplest approximation is to
neglect \textit{all} many-body effects and consider the one-electron problem
when self-energy (Fig. 1(b)) and screening (Fig. 1(c)) effects disappear 
leaving the standard [4] excitonic binding problem (Fig. 1 (a)) for a 
conduction band electron and a valence band hole interacting via the 
effective "one dimensional" Coulomb interaction. Even this zeroth order 
exciton problem for quantum wires is far from trivial, however, because one 
must include proper quantum confinement effects in the Coulomb interaction 
matrix elements appropriate for the specific quantum wire geometry of interest.
Not surprisingly a rather large theoretical literature [4,5] exists in 
treating this zeroth order one-electron quantum wire exciton problem, which 
is the effective dilute or zero density ($n\rightarrow 0$) limit of the 
many-body problem of interest to us. We include quantum wire 
confinement effects appropriate for a T-junction system in all the results
presented in this paper.

\begin{figure}
 \vbox to 5.9cm {\vss\hbox to 5.5cm
 {\hss\
   {\includegraphics{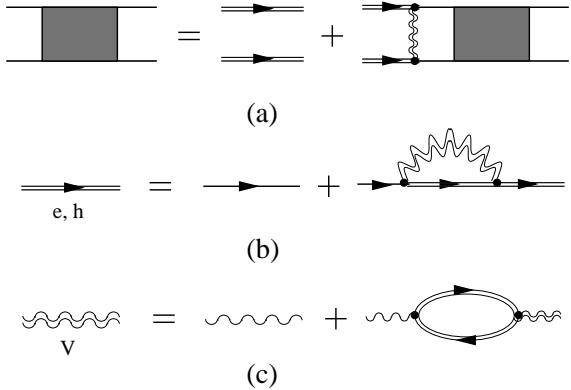}
   }
  \hss}
 }
\caption{
Many-body Feynman diagrams used in the theory with the single (double)
solid line representing the bare (dressed) electron or hole Green's function,
the single (double) wavy line representing the bare (dressed) Coulomb
interaction: (a) the excitonic Bethe-Salpeter equation; (b) the single-loop
self-energy (in the so-called GW approximation) defining the dressed Green's
function; (c) the RPA dressing of the Coulomb interaction (treated in the
plasmon-pole approximation in our calculation).
}
\label{Fig:1}
\end{figure}
In carrying out the full many-body dynamical calculation for the 
Bethe-Salpeter equation
we are forced to make some approximations. Our most sophisticated approximation
uses the fully frequency dependent dynamically screened electron-hole Coulomb 
interaction in the single plasmon-pole random phase approximation (Fig. 1(c)), 
which has been shown to 
be an excellent approximation [11] for 1D quantum wire dynamical screening.
It is essential to use the 
actual Coulomb interaction in solving this problem and the simplistic model 
interactions (such as the delta function zero range interaction used recently 
in Ref. 6) are not particularly meaningful from either a theoretical 
perspective or in understanding experimental data. In addition to our full 
dynamical screening theory (which is computationally extremely difficult) we 
have also carried out a number of simpler approximations 
(to be described below) in order to assess 
the quantitative contributions of various physical mechanisms to the 1D 
many-body exciton formation. For the self-energy correction we use the 
single-loop GW diagram shown in Fig. 1(b). Ward Identities then fix the 
vertex correction, entering Fig. 1(a), to be the appropriate ladder integral 
equation.

\begin{figure}

 \vbox to 6.5cm {\vss\hbox to 6cm
 {\hss\
   {\includegraphics{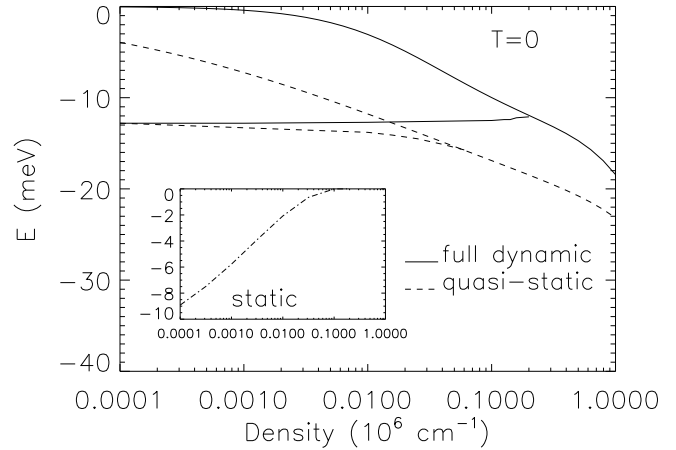}
   }
  \hss}
 }
\caption{
Separately calculated exciton energy and BGR (shown with 
respect to the free band gap as the energy zero) of the EHP 
as a function of photoexcitation carrier density. 
The heavy lines are the exciton energy 
and the light lines are the BGR
correction obtained from the self-energy diagram (Fig. 1(b) with 1(c)) 
neglecting the excitonic binding effect of Fig. 1(a). The solid and 
dashed lines are respectively the full dynamical screening and quasi-static 
approximations as described in the text. The Mott 
transition occurs at the density ($n_c$) where the heavy and the light lines 
cross indicating the exciton merging with the band continuum. The inset shows 
the exciton binding energy in the statically screened 1D Coulomb interaction 
neglecting BGR effects in an effective
single-exciton picture.
}
\label{Fig:2}
\end{figure}
Before solving the full Bethe-Salpeter equation, it is instructive to study 
the excitonic and EHP effects \textit{separately} by treating the influence
of the plasma on the excitonic states as a perturbation [10]. Using
an effective Hamiltonian derived from the Bethe-Salpeter equation, 
we can obtain the exciton energy by minimizing the energy expectation value
variationally through an 1s excitonic trial wave function.
The BGR is calculated by the GW approximation (Fig. 1(b)).
Note that the variational calculation is quantitatively valid only when the 
exciton-plasma hybridization is not particular important. 
In Fig. 2 we show
our calculated zero-temperature (variational) excitonic energy and BGR 
separately as a function of 1D electron-hole density. The full dynamical
screening solution is shown as the solid line and the 
\textit{quasi-static} screening approximations (described below) is shown as 
the dashed line. For the purpose of comparison we also show as an inset
in Fig. 2 
the purely \textit{one-electron} static screening result where the 
electron-hole interaction is modeled by the density dependent statically 
screened 1D Coulomb interaction, and all many-body effects (e.g. BGR) are 
ignored. The exciton binding energy 
shows a monotonic decrease ("blue-shift") in the inset (induced by 
static screening) as the exciton 
eventually merges with the band continuum with a Mott transition density 
$n_c\sim10^5$ cm$^{-1}$. The
quasi-static approximation [10], shown as dashed lines in Fig.2, involves 
making the screened exchange plus Coulomb hole approximation in
the self-energy diagrams neglecting the correlation hole 
effect. The simpler approximations (static screening and quasi-static)
are done in order to assess the importance of various terms 
in the full dynamical Bethe-Salpeter equation which is extremely difficult 
and computationally time-consuming to solve in the RPA dynamical 
screening approximation.

\begin{figure}

 \vbox to 8.5cm {\vss\hbox to 6cm
 {\hss\
   {\includegraphics{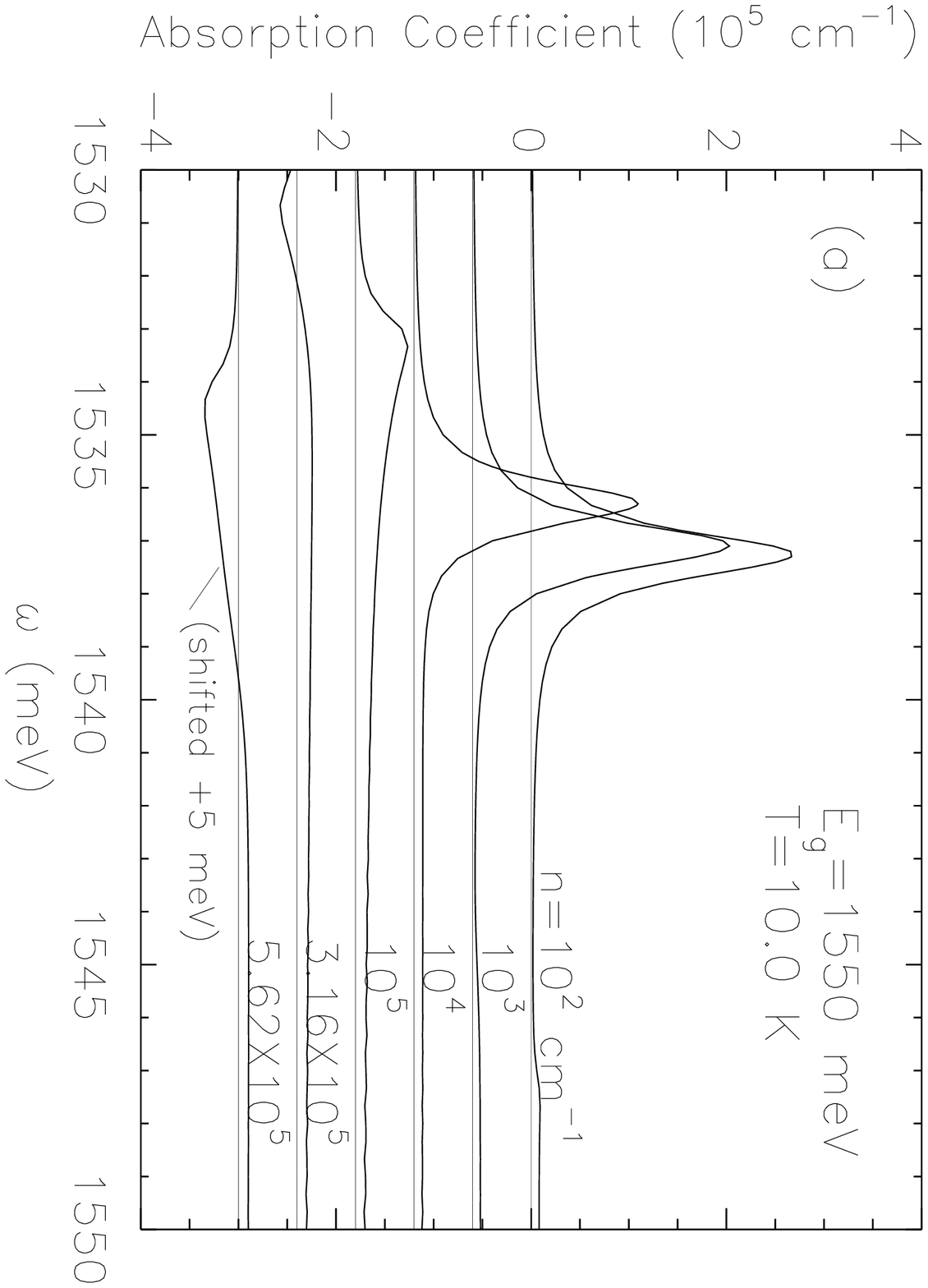}
   }
  \hss}
 }
 \vbox to 4.4cm {\vss\hbox to 6cm
 {\hss\
   {\includegraphics{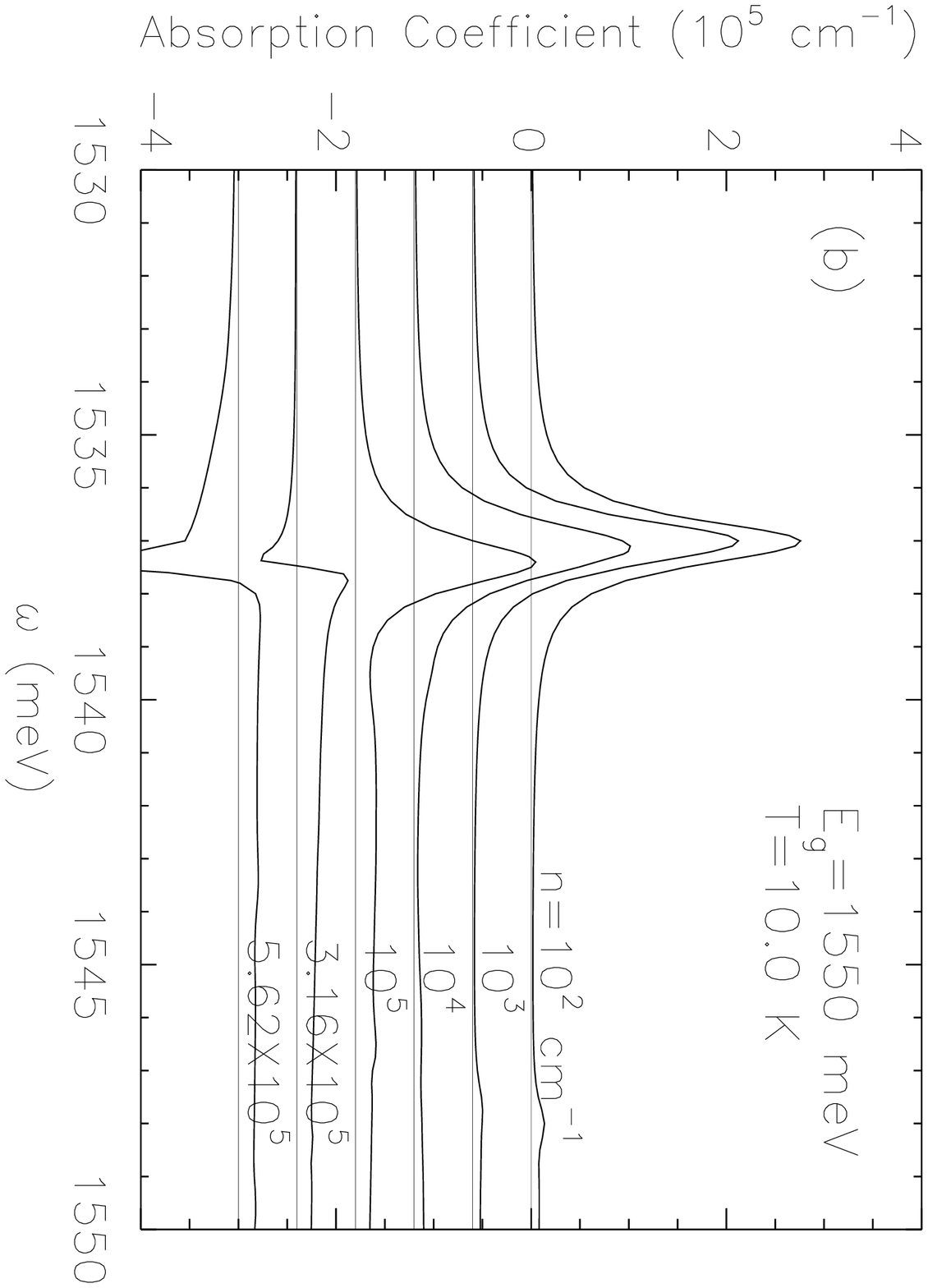}
   }
  \hss}
 }
\caption{
Calculated absorption and gain spectra for various photoexcitation densities
by solving the full Bethe-Salpeter equation: (a) the quasi-static
approximation, and (b) the full dynamical approximation. Negative
absorption indicates gain.
}
\label{Fig:3}
\end{figure}
The Mott transition may be thought of as the unbinding of the bound 
electron-hole pair in the exciton to a free electron and a free hole --- 
it is therefore effectively an interaction-induced insulator 
to metal transition which occurs as the exciton gas becomes an EHP 
at some high density ($n_c$). The statically screened single exciton behavior 
shown in the inset of Fig. 2 disagrees completely with the experimental 
finding of an approximately constant excitonic peak independent (at least in 
some finite range) of the free carrier density. We find [12] that this
large blue shift is not cancelled by the many-body self-energy effects within 
the same static screening approximation. Therefore, it is essential to consider 
the dynamical effects when one calculates the excitonic effects in quasi-1D
quantum wire systems. Inclusion of dynamical many-body  
effects, shown in the results in 
the main part of Fig. 2, qualitatively modifies the situation: (1) the 
effective many-body excitonic energy is almost the same in the low
density limit ($\sim13$ meV for $n<10^4$ cm$^{-1}$) in all the approximations; 
(2) for density 
between $10^4$ and $10^5$ cm$^{-1}$ the exciton energy has a few meV red-shift
in the quasi-static approximation and almost no shift (less than 0.5 meV 
blue-shift) in the dynamical screening approximation;
(3) the Mott transition density for the quasi-static approximation 
is about $10^5$ cm$^{-1}$, while it is about $3\times10^5$ cm$^{-1}$
for the dynamical theory; (4) below $n_c$
our variational solution corresponds to an excitonic wavefunction which is 
that of a bound electron-hole pair in the 1s hydrogenic state
with a radius of about 100-500 \AA\ [12], 
and this description is approximately valid with a constant (variational)
ground state energy upto $n_c$ [12]; (5) above $n_c$ the calculated effective
excitonic wave function is completely delocalized (with a very large radius)
and the EHP becomes the dominant state of the system; (5)
the quasi-static approximation, while being qualitatively valid, is quite poor
quantitatively compared with our dynamical screening approximation.

In Fig. 3, we show our calculated absorption and gain spectra 
by solving the full Bethe-Salpeter equation in the quasi-static 
and the dynamical screening approximations. The integral equation for the
two-particle Green's function (Fig. 1(a)) is solved by the matrix
inversion method with a singular kernel [10] which arises from the 
singularity of the Coulomb interaction.
The full dynamical screening approximation (which has never been solved in 
the literature before) has a multi-singular kernel
with multiple momentum-dependent singularities (poles of the integrand)
which arise from the many-body hybridization 
of photons, single particle excitations, and plasmons. This makes the usual
singularity-removal method ineffective. This fact forces us to use
a rather large matrix (about $1500\times1500$ in a Gaussian quadrature) in the 
matrix inversion [10] method
in order to get good overall accuracy.
We now discuss the important features of Fig. 3:
(1) There are generally two absorption peaks in the low density 
($n<10^4$ cm$^{-1}$) spectra, one is the exciton peak at 1537 meV and the other
one is the band edge peak at, for example, 1547.5 meV for $n=10^2$ cm$^{-1}$ 
in Fig. 3(b). The exciton peak has much 
larger oscillator strength than the band edge peak. 
(2) At low densities ($n<10^4$ cm$^{-1}$)
the exciton peak does not shift much ($\sim1537$ meV) with
increasing carrier density (in either approximation), 
indicating the effective constancy of the exciton energy; (3)
at higher densities, however, the quasi-static approximation produces a 
red-shift in the excitonic peak by a few meV, consistent with the result
shown in Fig. 2 which was obtained variationally. (4) Consistent with the 
variational energy shown in Fig. 2, the excitonic peak of the full dynamical
screening approximation is almost a constant (with only a 0.5 meV blue-shift) 
upto $n_c$. (5) Below the Mott density ($n_c\sim3\times 10^5$ cm$^{-1}$)
the oscillator strength of the excitons 
decreases rapidly as the carrier density increases in the quasi-static 
approximation; however, in the full dynamical theory the strength of the 
exciton peak remains almost a constant with increasing carrier density, 
indicating the interesting prospect of excitonic lasing in 1D quantum wires.
(6) In the dynamical screening approximation, considerable excitonic gain is
achieved for $n>n_c$ without any observable energy shift in the spectrum.
We find that at very high densities ($n>10^6$ cm$^{-1}$) the excitonic 
features in the absorption spectra are smeared out by the EHP continuum, 
and the BGR induced red-shift is observed. These very high density results 
will be presented elsewhere [12].

We note that our dynamical screening Bethe-Salpeter equation results are in 
excellent qualitative and quantitative agreement with the recent experimental 
findings [1,2]. In particular, the effective constancy of the exciton peak 
as a function of the photoexcited carrier density as well as the possibility 
of excitonic absorption and lasing well into the high density regime 
(even for $n>n_c\sim3\times 10^5$ cm$^{-1}$) turn out to be characteristic 
features of the full dynamical theory (but \textit{not} of the static and 
the quasi-static approximation). A full dynamical self-consistent theory as 
developed in this Letter is thus needed for an understanding of the recent 
experimental observations. We also note that in the recent literature the 
Mott density for 1D GaAs quantum wire systems has often been quoted as 
$n_c\sim8\times 10^5$ cm$^{-1}$ which is substantially higher than our full 
dynamical theory result, $n_c\sim3\times 10^5$ cm$^{-1}$. 
The higher value of the Mott density ($n_c\sim8\times 10^5$ cm$^{-1}$) 
follows from a simple estimate based on ground state energy comparison 
where one equates the calculated 
density-dependent BGR (the light solid line in Fig. 2) with the zero-density 
exciton energy ($\sim13$ meV in Fig. 2) --- as one can see from Fig. 2, the 
calculated BGR (the light solid line in Fig. 2) equals 13 meV, the 
zero-density exciton energy, around $n\sim8\times 10^5$ cm$^{-1}$. 
In the full interacting theory the Mott transition (the 
intersection of the light and the heavy solid lines in Fig. 2) moves to a 
lower density, $n_c\sim3\times 10^5$ cm$^{-1}$, which is also consistent 
with our full dynamical Bethe-Salpeter equation based calculation of the 
absorption/gain spectra shown in Fig. 3.

In summary, our main accomplishments reported in this Letter are the following:
(1) The \textit{first} fully dynamical theory of a photoexcited electron-hole 
system in semiconductors which treats self-energy, vertex corrections, and 
dynamical screening in a self-consistent scheme based on the GW self-energy 
and ladder-bubble vertex-polarization diagrams within a realistic Coulomb 
interaction-based Bethe-Salpeter theory; (2) a reasonable qualitative and 
quantitative agreement with the recent experimental observations of an 
effectively (photoexcitation density-independent) constant exciton peak, 
which in our fully dynamical theory arises from an approximate cancelation 
of self-energy and vertex corrections in the Bethe-Salpeter equation; (3) an 
effective 1D quantum wire Mott transition density of $n_c\sim3\times 10^5$ 
cm$^{-1}$ which is below earlier estimates based on less sophisticated 
approximations; (4) the concrete theoretical demonstration of the possibility 
of excitonic gain and lasing in 1D quantum wire structures in the 
density range of $n>3\times 10^5$ cm$^{-1}$ where considerable optical gain 
is achieved in our calculated absorption spectra. 

In conclusion, we have carried out the first fully dynamical many-body theory
for the photoexcited electron-hole plasma in 1D semiconductor quantum wires by 
solving the Bethe-Salpeter equation treating self-energy ("band gap 
renormalization") and vertex ("excitonic shift") corrections on an equal 
footing within the ladder-bubble-GW self-consistent conserving scheme. We 
find, consistent with a number of hitherto unexplained experimental 
observations [1-3], that the self-energy and the vertex corrections tend to 
cancel each other leading to an almost constant (in density) absorption/gain 
peak all the way to (and considerablely above) the Mott transition which occurs around a density of 
$n_c\sim3\times 10^5$ cm$^{-1}$ for 70 \AA\ wide T-quantum wires.

This work has been supported by the US-ONR and the US-ARO.


\end{document}